\documentstyle[12pt]{article}
\begin{document}
\title{Inelastic Rescattering in
 $B \rightarrow \pi \pi , K\bar{K}$ 
\footnote{Presented at the Cracow Epiphany Conference on $b$ Physics
and CP Violation, Cracow, Poland, January 5-7, 2001.} }
\author{
{P. \.Zenczykowski}$^a$\\
{\em H. Niewodnicza\'nski},
{\em Institute of Nuclear Physics}\\
{\em Radzikowskiego 152,
Krak\'ow, Poland}\\
}
\maketitle
\vspace*{-0.5cm}
\begin{abstract}
Inelastic rescattering effects in
$B$ decays into a pair $PP$ of pseudoscalar mesons 
($PP = \pi \pi$ or $K\bar{K}$) are discussed. 
It is assumed that $B$ meson decays through a short-distance tree-diagram
process into two jet-like states 
composed of low-mass resonances $M_1 M_2$
which rescatter into $PP$.
The rescattering of resonance pair $M_1 M_2$  into the final $PP$ state
is assumed to proceed through a Regge flavour exchange.
Since such processes 
constitute a fraction,
diminishing with increasing energy,
of total inelastic $PP$  scattering,
the inelastic rescattering contribution should die out for $m_B \to \infty$.
At $m_B = 5.2 ~GeV$, however, explicit estimates show that rescattering
corrections could be substantial, leading to long-distance
corrections to $B^0 \to K\bar{K}$, comparable to short-distance
penguin contributions.
\end{abstract}
\vfill
$^a$ E-mail:
zenczyko@iblis.ifj.edu.pl
\newpage

Studies of CP-violation in $B$ decays must involve final state
interaction (FSI) effects. Unfortunately, a reliable estimate of such
effects is very hard to achieve. In the analyses of $B \rightarrow PP$ decays
($P$ - pseudoscalar meson) only some intermediate states, believed to provide 
nonnegligible contributions, are usually taken into account.
Most often studies are restricted to the case of 
elastic or quasi-elastic rescattering.

{\em 1. Quasi-elastic rescattering}\\
Consider quasi-elastic rescattering in $B^0$ decays into $\pi ^+ \pi ^-$
and $\pi ^0 \pi ^0$.
The short-distance (SD) amplitudes are
\begin{eqnarray}
w(B^0 \to \pi ^+ \pi ^-) &= &-\frac{1}{\sqrt{2}}(T+P)\nonumber \\
\label{shdist}
w(B^0 \to \pi ^0 \pi ^0) &= &\frac{1}{2}P
\end{eqnarray}
when only 
the dominant 
 $T$ (tree) and $P$ (penguin) amplitudes
are taken into account (Fig.1).
Since isospin is a good symmetry of strong interactions, 
rescattering is diagonal in the basis of a definite total isospin. 
Consequently,  one expects that 
the amplitudes $W$ of definite final isospin $I$ 
are modified by Watson phases $\delta_I$ only:
\begin{eqnarray}
W(B\to (\pi \pi)_0)& = & -\frac{1}{\sqrt{3}}(T+\frac{3}{2}P) \exp (i\delta_0)
\nonumber \\
\label{isospin}
W(B\to (\pi \pi)_2)& = & -\frac{1}{\sqrt{6}}T \exp (i\delta_2)
\end{eqnarray}
The physical assumption that goes into Eq.(\ref{isospin})
is that the probabilities of SD decays are not changed by such a 
long-distance (LD) effect as rescattering.
When decay amplitude into a (eg.) $\pi ^+ \pi ^-$ state is extracted
from Eqs(\ref{isospin})
one obtains
\begin{equation}
\label{addition}
W(B^0 \to \pi ^+ \pi ^-) = - \frac{1}{\sqrt{2}} \exp (i\delta _2)
\left(
T+\frac{2}{3}T(\exp (\Delta )-1)+P \exp (i \Delta)\right)
\end{equation}
where $\Delta = \delta _0 -\delta _2$.
The appearance of the second term on the r.h.s. of 
Eq.(\ref{addition}) indicates that  in general final-state interactions
cannot be properly described if one multiplies short-distance tree and
penguin amplitudes in Eq.(\ref{shdist})
by two different phase parameters \cite{Wolf95,GW97}.
The only exception is when the two phases $\delta_0$ and $\delta_2$
are equal, which occurs if only elastic $\pi \pi$ rescattering is allowed:
the Pomeron does not distinguish between $\pi ^+ \pi ^-$ and $\pi ^0 \pi ^0$,
or $(\pi \pi )_0$ and $(\pi \pi )_2$.

{\em 2. Short-distance penguin and rescattering contributions}\\
The short-distance penguin amplitude $P$ is estimated at 20\%
of the tree contribution. It is of the order of $\lambda ^3$, 
where $\lambda \equiv
\sin \theta _C$ is one of the four parameters in the Wolfenstein's
parametrization of the Cabibbo-Kobayashi-Maskawa (CKM) matrix. 
Due to the dominance by the intermediate top quark, $P$
has the weak phase $\beta $ from the $V_{td}$ element of the CKM matrix.
 Although $\lambda$ is not small enough to
permit real distinction between terms differing by one factor of $\lambda$,
we observe that if Fig.1b is understood as a diagram representing
low-energy rescattering 
through charmed ($q=c$) or noncharmed ($q=u$) intermediate states, these
 processes are also of the order of $\lambda ^3$. For $q=c$, the amplitude
 of this "charming penguin" 
has weak phase $0$.
For $q = u$, the diagram represents quasi-elastic ($\pi \pi \to \pi \pi$)
and inelastic ($M_1M_2 \to \pi \pi$) rescattering. The corresponding
amplitude has weak phase $-\gamma$
of the $V_{ub}$ element of the CKM matrix.
With one of the goals of the present program of CP violation studies
being the determination of $CP$-violating phases $\alpha, \beta, \gamma $,
it is clearly very important to know which of the penguin-like admixtures
to tree diagrams is in fact dominant (if any) and, consequently, 
which weak phase will be probed by experiments.

{\em 3. Simple two-channel model} \\
Since a large part of the (time-reversal invariant) inelastic $\pi \pi$ 
scattering at energy $\sqrt{s}=m_B=5.2~GeV$ (or larger) 
goes into multiparticle final states composed of noncharmed mesons,
one may conjecture 
that inelastic rescattering processes shown in Fig.1b (with $q=u$) 
will be important.
A very simplified two-channel model of what should be expected 
of inelastic rescattering was presented in 
\cite{Don96etal,SuzWolf99}.
In this model there are two states: $|f_1\rangle =|\pi \pi \rangle $, and 
$|f_2\rangle $ representing "everything else" that $\pi \pi$ might
scatter into.
The most general $2\times 2$ unitary $S$ matrix is
\begin{equation}
\label{2by2}
S=\left[
\begin{array}{cc}
\cos 2\theta & i \sin 2\theta\\
i\sin2\theta & \cos 2\theta
\end{array}
\right]
\end{equation}
with the top left element describing $\pi \pi \to \pi \pi$ scattering.

If one accepts that 
final state interactions cannot modify the probability of the original SD
weak decay, it follows that vector ${\bf W}$ representing the 
FSI-corrected amplitudes is related to vector ${\bf w}$ of the original 
SD amplitudes through \cite{SuzWolf99}
\begin{equation}
\label{eq:S1/2}
{\bf W} = {\bf S}^{1/2} {\bf w}\approx ({\bf 1}+\frac{i}{2}{\bf T})~{\bf w}
\end{equation}
(where the last equality follows if ${\bf S}$ is close to ${\bf 1}$, ie. if
rescattering is assumed small).
The appearance of the square root of the ${\bf S}$ matrix is related
to Watson's theorem: 
 in the basis of ${\bf S}$-matrix eigenstates $|\nu \rangle $, 
the above equation reduces to 
$W_{\nu}=e^{i\delta_{\nu}}w_{\nu}$, ie. the
condition of unchanged probability 
($|W_{\nu}|=|w_{\nu}|$) admits Watson phases only.

With
\begin{equation}
\label{S12}
{\bf S}^{1/2}=
\left[
\begin{array}{cc}
\cos \theta & i \sin \theta \\
i \sin \theta & \cos \theta
\end{array}
\right]
\end{equation}
one immediately obtains
\begin{equation}
\label{2by2FSI}
W_1=w_1 \cos \theta  + i~ w_2 \sin \theta  \equiv |W_1| e^{i \delta _1}
\end{equation}
Since from $\pi N$ data extrapolated to $\pi \pi$ one gets
$\cos 2\theta \approx 0.6-0.7$,
one finds
\begin{equation}
\label{2by2delta1}
\tan \delta _1 = \frac{w_1}{w_2} \tan \theta \approx 0.45 \frac{w_1}{w_2}
\end{equation}
If B meson decays with similar strengths into the $|\pi \pi \rangle $ 
 and 
 $|f_2\rangle$ states, one calculates $\delta _1 \approx 20^o-25^o$.
In principle, therefore, the effect of inelastic rescattering may be large.

{\em 4. Inelastic rescattering at high energy}\\
In the previous example all states produced in the first stage of the decay
could be rescattered into $\pi \pi$. This is an overpessimistic situation. 
Consider a little more general case with three states 
$|f_1\rangle \equiv |\pi \pi\rangle $, $|f_2\rangle$, and $|f_3\rangle $,
 into which
$\pi \pi$ scattering may go. It may happen that $B$ meson does not decay
into state $|f_3\rangle$. If $\pi \pi$ inelastic scattering is dominated
by $|\pi \pi \rangle \to |f_3\rangle$ transitions with $|\pi \pi \rangle 
\to |f_2\rangle $ small or
negligible, then the total inelastic rescattering contribution corresponds to
$|B\rangle \to |f_2\rangle \to |\pi \pi \rangle$ and must be fairly weak. 
In such a case neglecting
final state interaction altogether might be a good approximation:
elastic FSIs result in the multiplication of SD amplitudes by the same 
phase for all isospin- (or SU(3)-symmetry-) related final states.

In order to know whether neglecting inelastic rescattering is or is not
justified, we have to estimate the
contribution from  states of type $|f_2 \rangle $ (ie. those
into which $B$ decays)
in the unitarity relation for the $l=0$ partial wave in $\pi \pi$ 
scattering
\begin{equation}
\label{unitarityf2f3}
|\langle \pi \pi |S|\pi \pi \rangle|^2+
\sum_{k}|\langle\pi \pi |S|f_k\rangle|^2=1.
\end{equation}

Inelastic production of particles in high energy $\pi \pi$ scattering 
is expected to proceed after the projectiles have exchanged
a gluon. Colour separation is prevented by the production of many
$q\bar{q}$ pairs whose number increases with energy.
Colour singlet quark-antiquark pairs  materialize later as resonances whose
momenta  
are approximately parallel to the axis of  $\pi \pi$ collision
in c.m.s.
Thus, a typical inelastic process may be represented by the diagram
shown in Fig.2a, in which the ordering of quark lines from top to bottom
may be roughly correlated with the ordering 
in rapidity of produced resonances.
Any transfer of flavour quantum numbers over large distances in rapidity
is suppressed.
This corresponds to suppression of flavour-exchange in quasi-two-body
production processes $\pi \pi \to M_1M_2$ at high energies
(the time-reversed process of this type is shown on the r.h.s. 
of the diagram in Fig.1b with $q=u$).
It is through the presence of many free quark-antiquark pairs of 
{\em unconstrained} flavour
that the combinatorial factors beat suppressions of flavour transfer and,
through the production of more and more resonances with increasing energy, 
they maintain approximately constant inelastic cross section.

As a particular realisation of these ideas we may consider the multiRegge
model of resonance production discussed over twenty years ago \cite{Chanrev}.
In this model inelastic $\pi \pi$ collisions produce a variable 
number of resonances ordered along the rapidity axis with Regge exchanges
in between the resonances. As energy ($m_B$ in our case) increases, 
the average number of Regge exchanges in question and the average number 
of resonances produced must increase, so that no flavour exchange over 
a large distance in rapidity space occurs. 
Thus, at high energy one expects the production of two-resonance
states $|f_2\rangle$ to die out. The same will happen to the production
of three-resonance states, and so on,
while the majority of the inelastic cross section will shift to the
production of $k$-resonance states  with average $k$
increasing with energy. This situation might be described by replacing
$|f_3\rangle $ with a set of $k$-resonance states $|f_k\rangle $.
(In fact, each of states $|f_k\rangle $
should be labelled with an additional index
to distinguish between states consisting of different $k$ 
resonances.)
A first thought is that states $|f_k\rangle $ for $k \ge 3$ 
cannot contribute to rescattering in B decays.
However, since resonances may decay to lighter ones,
interference terms may occur. We will argue, however, that
such terms 
should be small and vanishing at high energy.

As an example, let us consider the case when $m_B $ is large and the 
heavy resonance $M_1$ from state
$|f_2\rangle $ 
decays into three lighter resonances,
which later rescatter through $|f_4\rangle \to |\pi \pi \rangle $
(the argument is better when the number of resonances
into which resonance $M_1$ decays is larger, ie. when $m_{M_1}$ 
and $m_B$ are larger).
The rescattering part of the process (the diagram on the r.h.s. of Fig.2b) 
cannot contain quark lines connecting
outgoing pions: such processes are suppressed at high energies.
Thus, there must be quark-antiquark pairs produced out of the
vacuum (eg. the line connecting the first and the third topmost resonances).
When one takes
the square of the amplitude corresponding to 
the diagram on the r.h.s. of Fig.2b to evaluate its contribution
to inelastic $\pi \pi$ scattering, one obtains amplitude
represented by a diagram with free quark loops of unconstrained flavour.
These loops
lead to a large combinatorial factor mentioned earlier.
However, when the left- and right-hand sides of the diagram of Fig.2b
are combined to represent the decay followed by rescattering, 
the combinatorial factor 
is dramatically reduced: there are no quark lines of unconstrained flavour.
 Within the language of ref.\cite{Chanrev}, the resulting combined diagrams
 are even more suppressed:
namely those parts of a diagram 
in which quark lines of an intermediate resonance "cross" 
(the second resonance from the top in Fig.2b)  
represent cancellations between resonances of opposite C-parities.
Thus, only low-mass resonances $M_1$ and $M_2$ will contribute
to the rescattering.
Since with increasing energy inelastic $\pi \pi $ scattering is
dominated by channels with an ever-increasing number of produced resonances,
it follows that
inelastic rescattering in $B$ decays should die out at $m_B \to \infty$.
It is interesting to note that the
dissappearing of rescattering effects at $m_B \to \infty $ was found also
in a different framework \cite{SachrajdaEpiph}.

The mismatch between the left- and right-hand sides of the diagram in Fig.2b
should be attributed to the difference in the original separation of colour: 
receding 
quarks of the decaying $M_1$ resonance stretch a $3-\bar{3}$ colour string,
while in the inelastic rescattering the dominant part comes from octet
separation. The latter requires an  
exchange (diminishing with increasing energy)
of a quark to reduce it to
$3-\bar{3}$. Since vanishing of inelastic rescattering at $m_B \to \infty$
is related to colour mismatch, this vanishing should be present in all models.

{\em 5. Inelastic rescattering at $m_B=5.2~GeV$}\\
Although inelastic rescattering is expected to die out at high 
energy, one may ask if the value of $m_B=5.2~GeV$ is sufficiently high 
to neglect it.
In order to make some estimates, we turn to multiRegge models of multiparticle
production processes \cite{Chanrev}.
In these models the distribution of the number of produced resonances 
was evaluated as a function of energy. It turns out \cite{CPTN75}
that in the $\pi \pi $ scattering at $\sqrt{s}=5.2~GeV$, final states composed
of just two resonances occur in 50\% of cases, while three-resonance
states show up in 35\% cases, and so on.
Only at much higher $\sqrt{s}$ of the order of $20-30~GeV$ and more, 
the production
of three and more resonances becomes dominant.
Thus, inelastic rescattering is not negligible
at $m_B=5.2~GeV$, and one has to estimate it explicitly.

In order to do that, we assume that 
the average amplitude for the $PP \to M_1M_2$ process is approximately
equal to the flavour-exchange Regge amplitude in $PP \to PP$.
From $\pi N$ data extrapolated to $\pi \pi$ scattering one can estimate
that at $\sqrt{s}=5.2~GeV$,
 the $l=0$ partial wave projection of the latter leads to
\begin{equation}
\label{ReggeonPP}
|\langle PP|S_{Reggeon}|PP\rangle|^2=0.025
\end{equation}
for a typical state $|PP\rangle $ 
averaged over different SU(3)-flavour representations of the $PP$ system
\cite{PZPRD2001,Zen99}.
On the other hand, Pomeron exchange yields
\begin{equation}
\label{Pomeron}
\langle PP |S_{Pomeron}| PP \rangle = 
\langle \pi \pi |S_{Pomeron}| \pi \pi \rangle = \cos 2\theta
\approx 0.65
\end{equation}
If one assumes further that all $\langle PP|S|f_k\rangle $ amplitudes
are approximately equal to $\langle PP|S_{Reggeon}|PP\rangle $ 
(for average state of $|f_2\rangle $ type this assumption is
in agreement with phenomenological analyses \cite{PZPRD2001,HRR73}),
one obtains
from Eq.(\ref{unitarityf2f3}):
\begin{equation}
|\langle PP |S_{Pomeron}|PP\rangle|^2 +
n_{tot}|\langle PP |S_{Reggeon}|PP\rangle|^2 = 1
\end{equation}
where $n_{tot}$ denotes the number of average channels.
Using Eqs(\ref{Pomeron},\ref{ReggeonPP}) one estimates that
\begin{equation}
\label{ntot}
n_{tot} \approx 25.
\end{equation}
or more (\cite{PZPRD2001}).
With 50\% of the total cross section at $\sqrt{s}=5.2~GeV$ being
estimated as due to the production of two-resonance states, one finds
that the number of two-resonance states is of the order of
\begin{equation}
\label{n2}
n_2 \approx 12
\end{equation}

One may also try
to directly estimate the number of $M_1M_2$ states produced by the 
short-distance decay mechanism. The SD decay amplitude
 is dominated by the tree-diagram contribution.
The $M_1M_2$ states produced through this process include both states 
composed of light pseudoscalar mesons ($\pi \pi $ in the simplest case)
as well as various resonances. 
From the known $b$-quark SD decay probability 
\begin{equation}
\label{bdecay}
v(q^2)=2 (1-q^2/m^2_b)^2 (1+2 q^2/m^2_b)
\end{equation}
one can see that for $m_B \approx 5.2~GeV$
even quite massive resonances $M_1$  (3-4~GeV) could be formed.
The $M_2$ spectrum extends to $m_2 \approx 2.0~{\rm or}~ 2.5~GeV$.
The average $m_1$ mass is of the order of $2.5~GeV$, while $m_2$ is
a little smaller, around $1.4~GeV$. 
The number of $M_1M_2$ states
produced in the above range of masses 
may be estimated on the basis of the ISGW2 approach \cite{ISGW2}.
One obtains \cite{PZPRD2001} around 10-20 different $M_1M_2$ states,
in agreement with Eq.(\ref{n2}).

In order to perform more realistic estimates of rescattering effects 
and the way they might affect CP violation phenomena, it
is important to know the phases of $M_1M_2 \to PP $ amplitudes.
These may be estimated by analysing Regge behaviour of such
amplitudes in dual models \cite{Chanrev}.
One finds that for the FSI
amplitudes of the r.h.s. of Fig.1b the phase factor arises from
\begin{equation}
\label{rotating}
\left(
-\frac{s}{m^2_1m^2_2}
\right)^{\alpha (t)}
\end{equation}
where $\alpha (t)$ is the Regge trajectory.
If the exchanged Reggeon just interchanges two quarks in the $s$-channel,
the phase is real. 
After projecting Regge amplitudes onto the $l=0$ partial wave, 
the dependence on resonance masses seen in Eq.(\ref{rotating})
gives rise to a phase factor depending on the product $(m_1m_2)^2$.
For low $M_1,M_2$ masses, wherefrom
the dominant part of rescattering is expected to come, this phase
turns out to range from about $-40^o$ at $(m_1m_2)^2=1~GeV^4$ to
$+20^o$ at $(m_1m_2)\approx 7~GeV^4$, where Regge approximation
starts to become questionable (approximation of Eq.(\ref{rotating})
requires values of $s/(m_1m_2)^2$ larger than $4~GeV^{-2}$ or so).

Contributions from rescattering through different $M_1M_2$ states add up
according to the distribution of $(m_1m_2)^2$ produced in SD decay.
Using Eq.(\ref{eq:S1/2}) and
an appropriately chosen $(m_1m_2)^2$ distribution, the combined 
effect of rescattering through all intermediate inelastic states can be
estimated from
\begin{equation}
\label{combined}
\langle PP |W|B\rangle = \langle PP |1-\frac{1}{2} {\rm Im}~T|PP\rangle
\langle PP|w|B\rangle +
\frac{i}{2}\sum_{j=1}^{n_2}\langle PP|T|f_{2,j}\rangle
\langle f_{2,j}|w|B\rangle
\end{equation}
where the sum over $j$ from $1$ to $n_2$
in the second term corresponds to the summation
over different values of $(m_1m_2)^2$ starting from its smallest value and
according to its distribution.
The first term in Eq.(\ref{combined}) describes SD decay amplitude 
with corrections due to elastic
rescattering.
Since the phases of amplitudes $\langle PP|T|f_{2,j}\rangle$ are not far
from zero one expects that the phase of the
total contribution from inelastic rescattering
in B decays is around $90^o$.

Numerical estimates of rescattering in $B \to \pi \pi, K \bar{K}$ decays,
based on Eq.(\ref{combined}), lead to significant corrections to SD
estimates (see \cite{PZPRD2001}). As an example,
in Fig.3 the dependence on the number $n$ of  inelastic two-resonance
channels included
(ie. when $n_2$ in Eq.(\ref{combined}) is replaced by variable $n$)
is shown for the absolute size 
of the (zero isospin) $W((K\bar{K})_0)$ amplitude.
The amplitude is given in the units of the initial input tree
amplitude $T$.
In pure short-distance approach,
 this amplitude 
is equal to $W((K\bar{K})_0) = P_{SD}/2 \approx 0.1~ T$
since the relevant
penguin amplitude is usually estimated at $P_{SD}\approx 0.2~ T$.
As shown in Fig.3, the rescattering-induced contribution to $W((K\bar{K})_0)$
becomes equal to the SD contribution already at $n \approx 5$.

One has to conclude that although
for $m_B \to \infty$ one may expect that
the effects of inelastic rescattering in $B$ decays vanish,
for the physical value of $m_B=5.2~GeV$ these effects should be
still important. Consequently, unless some unknown reasons justifying the 
neglect of rescattering exist, the
extraction of CP-violating parameters from nonleptonic $B$ decays
will be most probably very difficult. One may still hope, however, 
that when enough
data on many different $B$ decay channels are accumulated, 
one will be able to remove the effects of FSIs, thus reaching
the parameters of the quark level.

\newpage
\noindent
FIGURE  CAPTIONS

\noindent
Fig.1\\
Dominant diagrams for $B$ decay:
(a) tree  $T$ and (b) penguin $P$ 

\noindent
Fig.2\\
(a) Multiparticle production at high energy\\
(b) Suppression mechanism for $B \to |f_4\rangle \to PP$

\noindent
Fig.3\\
Dependence of FSI-induced effects on the number of intermediate
channels included for $B \to (K\bar{K})_{I=0}$ decays
\\
\\


\begin{thebibliography}{99}
\bibitem{Wolf95} L. Wolfenstein, Phys. Rev. D52, 537 (1995).
\bibitem{GW97} J.-M. G\'{e}rard, J. Weyers, Eur. Phys. J. C7, 1 (1999).
\bibitem{Don96etal} J. F. Donoghue et al., Phys. Rev. Lett. 77, 2178 (1996).
\bibitem{SuzWolf99} M. Suzuki and L. Wolfenstein, Phys. Rev. D60:074019 (1999).
\bibitem{Chanrev} Hong-Mo Chan, Sheung Tsun Tsou "Dual Unitarization: a New
Approach to Hadron Reactions", RL-76-080, Sep.1976, published in
Bielefeld Sum. Inst. 1976, 83; G. F. Chew, C. Rosenzweig, Phys. Rept. 41,
263 (1978).
\bibitem{SachrajdaEpiph} C. Sachrajda, talk at this conference
and private communication. 
\bibitem{CPTN75} Hong-Mo Chan, J. E. Paton, Sheung Tsun Tsou, Sing Wai Ng,
Nucl. Phys. B92, 13 (1975); 
Hong-Mo Chan, J. E. Paton, Sheung Tsun Tsou, Nucl. Phys. B86, 479
(1975).
\bibitem{PZPRD2001} P. \.Zenczykowski, Phys.Rev.D63, 014016 (2001).
\bibitem{Zen99} P. \.Zenczykowski, Phys. Lett. B460, 390 (1999).
\bibitem{HRR73} P. Hoyer, R. G. Roberts, and D. P. Roy, Nucl. Phys. B56,
173 (1973)
\bibitem{ISGW2} D. Scora and N. Isgur, Phys. Rev. D52, 2783 (1995).

\end{thebibliography}
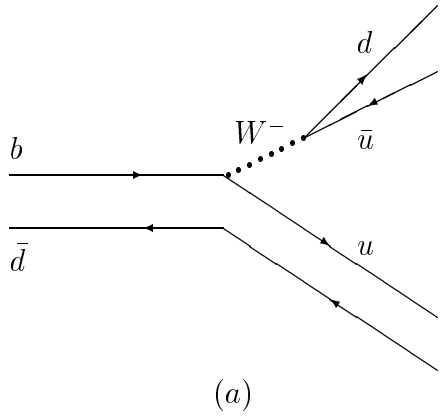
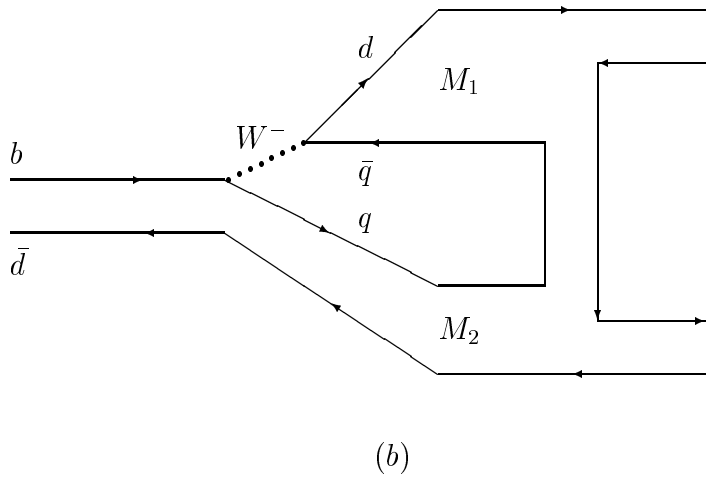
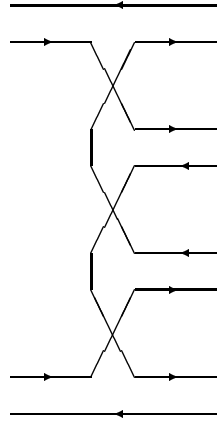
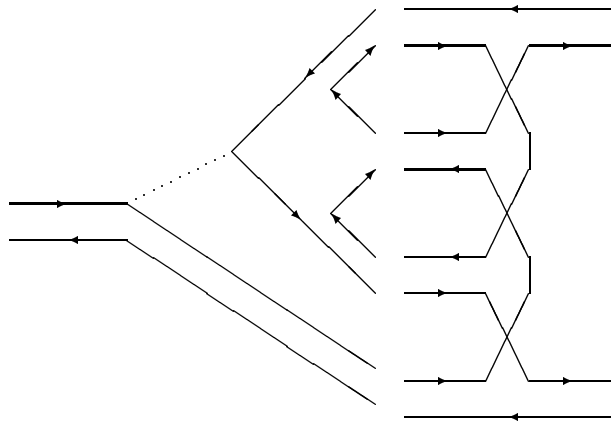
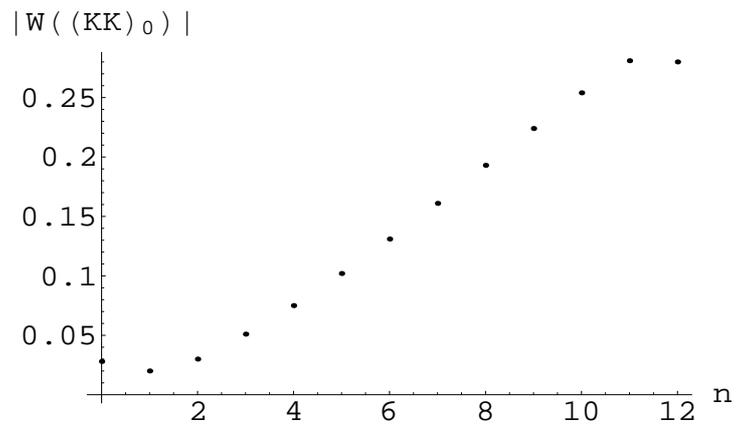
\end{document}